\documentclass[11pt]{article}

\usepackage[a4paper,margin=25mm]{geometry}
\usepackage[T1]{fontenc}
\usepackage{lmodern}
\usepackage{microtype}
\usepackage{amsmath,amssymb}
\usepackage{booktabs,array,tabularx,longtable}
\usepackage{graphicx}
\usepackage{xcolor}
\usepackage{algorithm}
\usepackage{algpseudocode}
\usepackage{float}
\usepackage{natbib}
\usepackage{placeins}
\usepackage{hyperref}
\usepackage[nameinlink,noabbrev]{cleveref}

\makeatletter
\providecommand{\theHALG@line}
  {\thealgorithm.\arabic{ALG@line}}
\makeatother

\hypersetup{
  colorlinks=true,
  linkcolor=blue!45!black,
  citecolor=green!35!black,
  urlcolor=blue!55!black
}

\newcolumntype{Y}{>{\raggedright\arraybackslash}X}

\title{VPID: An Integrated Framework for Vulnerability Prioritization
and Intrusion Detection in Enterprise Networks}

\author{%
\begin{minipage}[t]{0.31\textwidth}
  \centering
  Xuanren Chen\\
  {\small Hainan University}\\
  {\small Haikou, China}\\
  {\small arthurpz@163.com}
\end{minipage}
\hfill
\begin{minipage}[t]{0.31\textwidth}
  \centering
  Xin Wang\\
  {\small Hainan University}\\
  {\small Haikou, China}\\
  {\small 24210839000020@hainanu.edu.cn}
\end{minipage}
\hfill
\begin{minipage}[t]{0.31\textwidth}
  \centering
  Xiaoqi Li\\
  {\small Hainan University}\\
  {\small Haikou, China}\\
  {\small csxqli@ieee.org}
\end{minipage}
}

\date{}

\begin{document}

\maketitle

\begin{abstract}
Small enterprises face increasingly serious threats to their internal networks but often lack the financial resources, computing capacity, and specialist staff required to deploy resource intensive security platforms. This paper designs and implements VPID, a lightweight framework for vulnerability prioritization and intrusion detection that consists of two principal modules: controlled vulnerability validation and intelligent intrusion defense. The first module uses OpenVAS for asset mapping and vulnerability identification, applies a decision tree to prioritize vulnerabilities, and employs a rule engine to generate targeted validation payloads. The second module captures network traffic using Scapy, analyzes it through a detection pipeline that combines a decision tree with multinomial Naive Bayes, verifies traffic assessed as high risk using Snort rules, and performs blocking and alerting through iptables. The evaluation uses 550,000 network flow samples containing normal and attack traffic for detector training, together with 15,000 labeled vulnerability records. On the vulnerability ranking test set, the decision tree achieves a precision of 91.8
\%, a recall of 89.5\%, and an F1 score of 90.6\%. On an independent test set containing 55,000 traffic samples, the combined detection pipeline achieves a precision of 94.5\%, a recall of 88.3\%, and an F1 score of 91.3\%, while maintaining a false positive rate below 1.5\%.

\end{abstract}


\section{Introduction}\label{sec:introduction}

Small enterprises often operate with limited security budgets, modest computing resources, and few specialist security staff. These constraints create two closely related requirements. First, vulnerabilities must be prioritized so that limited remediation resources are directed toward the weaknesses that pose the greatest practical risk. Second, exploitation attempts must be detected and contained using mechanisms that can run on low-end hardware with limited manual intervention. A practical security solution for this environment should therefore be lightweight, interpretable, and capable of connecting vulnerability assessment with intrusion detection and response~\citep{li2026systematic}.Related security studies have also examined threats in emerging blockchain and decentralized-application environments from cryptographic, implementation, and application-level perspectives~\citep{yang2025multi,zhou2025blockchain,gao2025implementation}.

CVSS provides a standardized framework for measuring vulnerability severity \cite{first2019spec,first2019guide}. However, severity alone does not determine remediation priority. Previous studies have shown that practical prioritization also depends on factors such as asset value, network exposure, exploitability, temporal characteristics~\citep{jacobs2023enhancing}, and organizational context \cite{fruhwirth2009improving,jung2022cavp,le2022survey,jacobs2021exploit}. These findings motivate a context-aware prioritization method that considers both vulnerability severity and deployment-specific risk factorss~\citep{ullman2024enhancing,jiang2025survey,wang2024smart}.

For intrusion detection, Bayesian event classification has been used to reduce false alarms \cite{kruegel2003bayesian,li2025interaction}. Comparative studies have evaluated Naive Bayes and decision trees using intrusion detection datasets \cite{benamor2004naive}, while hidden Naive Bayes has been applied to attack detection involving multiple classes \cite{koc2012hidden}. Feature reduction and lightweight classification have also been widely investigated as approaches to improving detection efficiency \cite{mukherjee2012intrusion,kikissagbe2024review,zhang2025lightweight,abdulganiyu2023systematic}. Collectively, these studies demonstrate the potential of efficient and interpretable models in environments with limited resources~\citep{azimjonov2024designing}. However, detection performance at the model level does not by itself provide an operational workflow that connects traffic acquisition~\citep{momand2023systematic,pinto2023survey}, model inference, rule verification, and response enforcement~\citep{zhang2025penetration}. To address this gap, we design and implement VPID, a lightweight vulnerability prioritization and intrusion detection framework for resource-constrained small-enterprise networks. VPID integrates vulnerability discovery, context-aware prioritization, controlled validation, traffic detection, rule verification, and automated response through a unified processing pipeline~\citep{zhu2024sybil}. VPID combines mature open source security components with interpretable machine learning models through a unified processing pipeline. The main contributions are as follows:

\begin{enumerate}
  \renewcommand{\labelenumi}{(\arabic{enumi})}

  \item We propose VPID, a unified framework for resource-constrained small-enterprise networks. It connects OpenVAS-based vulnerability discovery with context-aware prioritization, controlled validation, traffic detection, and automated response.

  \item The vulnerability-prioritization method combines CVSS scores with exploitability, asset and business value, network exposure, patch status, exploitation history, and confidence. A CART model produces interpretable priority labels and comprehensive risk scores.

  \item For intrusion detection and response, we develop a two-stage pipeline that integrates decision-tree pre-screening, multinomial Naive Bayes classification, Snort-based verification, and iptables enforcement. This design balances detection performance, interpretability, and deployment cost.

  \item The VPID framework is implemented as four coordinated Docker services and evaluated using 15,000 labeled vulnerability records, 550,000 traffic-training samples, and an independent test set of 55,000 traffic samples. The vulnerability-prioritization model achieves 91.8\% precision, while the intrusion detector achieves 94.5\% precision with a false-positive rate below 1.5\%.
\end{enumerate}
\section{Technical Background}\label{sec:background}

\subsection{Vulnerability scanning}\label{subsec:scanning}

\begin{figure}[H]
  \centering
  \includegraphics[width=0.86\textwidth]{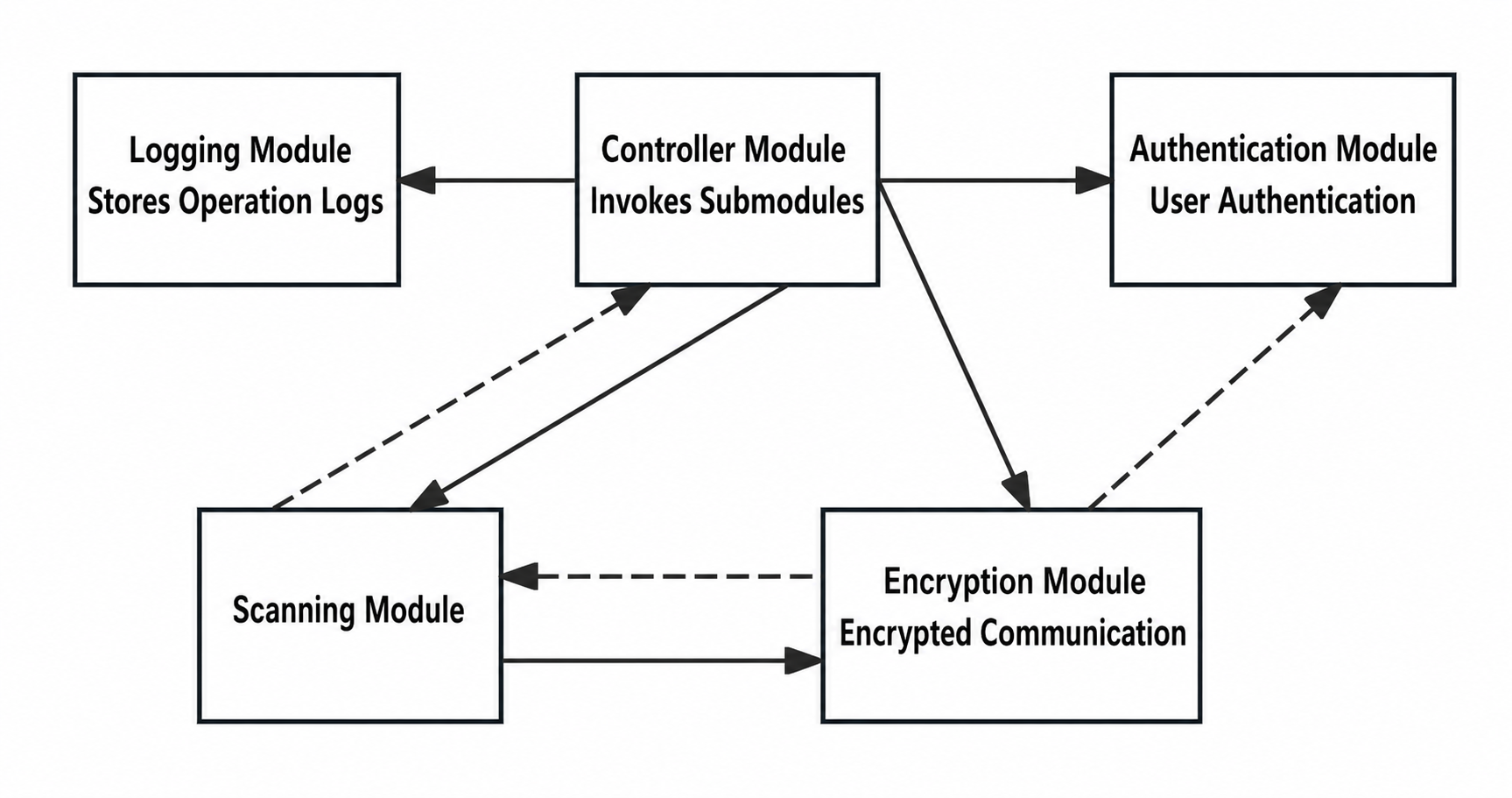}
  \caption{Supporting-module workflow for OpenVAS-based vulnerability scanning.}
  \label{fig:openvas-workflow}
\end{figure}

OpenVAS is an open source vulnerability scanner maintained within the Greenbone ecosystem. It executes an actively maintained collection of vulnerability tests against configured hosts and network services, enabling the identification of software weaknesses, exposed services, and insecure configurations. Its management interfaces support target definition, scan policy configuration, task scheduling, execution status monitoring, and structured report retrieval \cite{greenbone2024architecture,greenbone2024openvas}. The resulting reports contain information such as affected hosts, service ports, vulnerability identifiers, severity scores, technical descriptions, and remediation recommendations, allowing external applications to automate asset discovery and vulnerability collection while using OpenVAS as a standardized data source for subsequent risk analysis. Within VPID, a central controller uses these interfaces to coordinate the scanning process together with authentication, encrypted communication, and operational logging. It submits scanning targets and policies, monitors task execution, retrieves completed reports, and forwards the collected records to the preprocessing and vulnerability prioritization stages. Access to scanning operations is restricted by the authentication module, while encrypted communication protects data exchanged among system components. The logging module records scan requests, execution states, returned results, and processing errors to provide traceability and support subsequent auditing. The interactions among these supporting modules are shown in \cref{fig:openvas-workflow}.

CVSS v3.1 organizes vulnerability characteristics into Base, Temporal, and Environmental metric groups \cite{first2019spec,first2019guide}. The resulting score represents the technical severity of a vulnerability rather than its complete operational risk. The proposed system therefore treats the CVSS score as one input to vulnerability prioritization and combines it with information specific to each deployment during subsequent processing. The qualitative severity bands shown in \cref{tab:cvss-bands} follow CVSS v3.1, whereas the associated remediation targets are configurable operational policies implemented by the prototype and do not form part of the CVSS standard.

\begin{table}[htbp]
\centering
\caption{CVSS v3.1 severity bands and default remediation targets used by the prototype.}
\label{tab:cvss-bands}
\begin{tabular}{@{}lll@{}}
\toprule
CVSS score & Severity & Default remediation target \\
\midrule
9.0--10.0 & Critical & Within 24 h \\
7.0--8.9  & High     & Within 72 h \\
4.0--6.9  & Medium   & Within 30 days \\
0.1--3.9  & Low      & Quarterly review \\
0.0       & None     & No remediation target \\
\bottomrule
\end{tabular}
\end{table}

\subsection{Naive Bayes}\label{subsec:nb}

The intrusion-detection module uses multinomial Naive Bayes because it is computationally efficient for high-dimensional and sparse text representations. Let $X=(x_1,x_2,\ldots,x_n)$ be a nonnegative feature vector and let $C_k$ denote class $k$. According to Bayes' rule,

\begin{equation}\label{eq:bayes}
P(C_k\mid X)
=
\frac{P(X\mid C_k)P(C_k)}
     {P(X)}.
\end{equation}

Under the multinomial model, the class-conditional likelihood used for prediction is proportional to

\begin{equation}\label{eq:nb-independent}
P(X\mid C_k)
\propto
\prod_{i=1}^{n}\theta_{ki}^{x_i},
\end{equation}

where $\theta_{ki}$ is the smoothed conditional probability of feature $i$ in class $C_k$. To avoid numerical underflow, classification is performed in log space,

\begin{equation}\label{eq:nb-decision}
\hat{y}
=
\arg\max_{C_k}
\left[
\log P(C_k)
+
\sum_{i=1}^{n}x_i\log\theta_{ki}
\right].
\end{equation}

The multinomial and conditional-independence assumptions are approximations for correlated traffic features. Nevertheless, the model remains efficient for sparse representations and produces interpretable class scores that can be combined with other detection evidence.

\subsection{Decision trees}\label{subsec:trees}

\begin{figure}[H] \centering \includegraphics[width=0.7\linewidth]{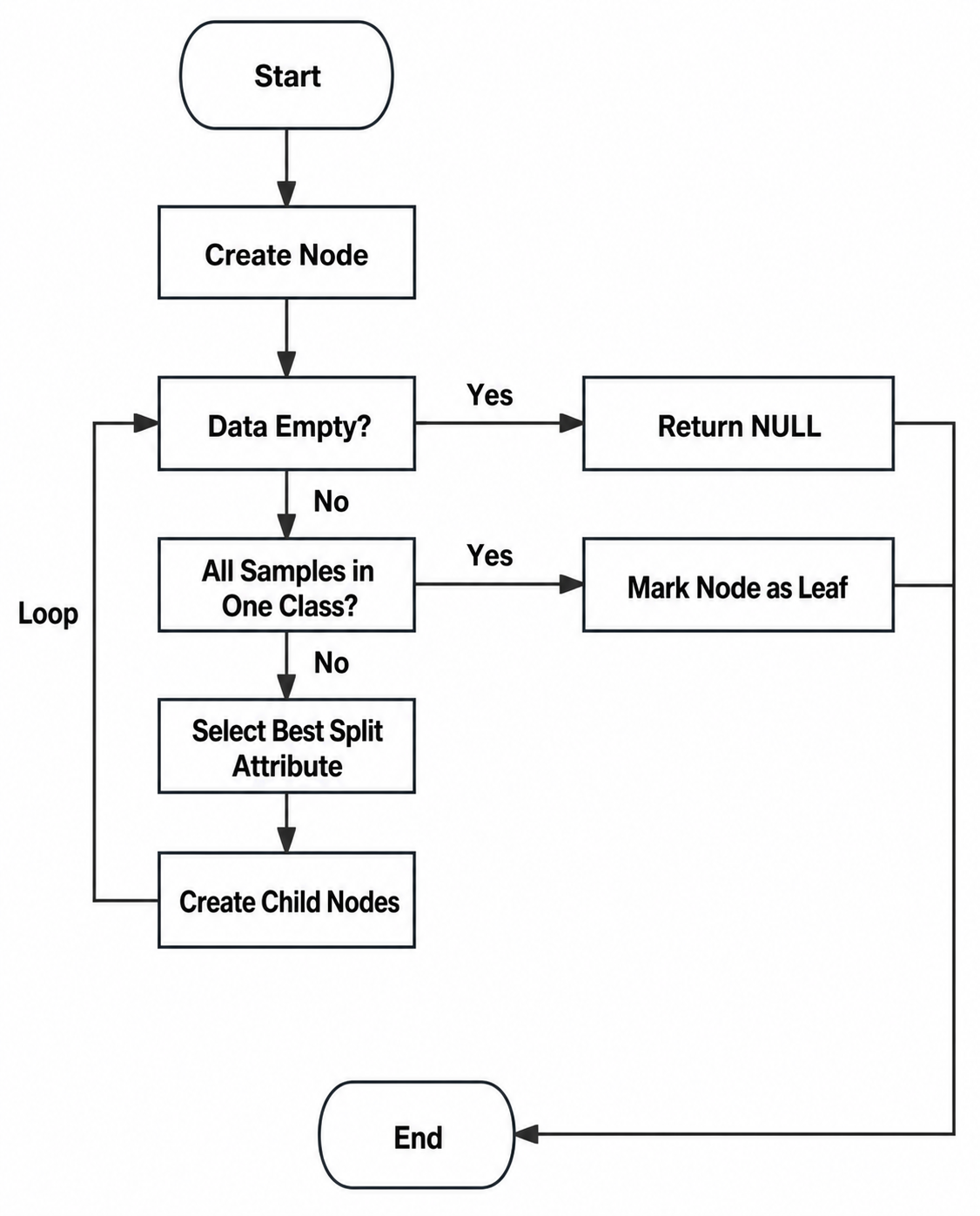} \caption{Decision-tree construction workflow.} \label{fig:tree-workflow} \end{figure}

Decision trees construct hierarchical classification models by recursively partitioning a dataset into increasingly homogeneous subsets \cite{breiman1984cart,quinlan1986induction}. At each internal node, the learning algorithm evaluates candidate features and split points and selects the split that produces the greatest reduction in class impurity. The same procedure is then applied to each child subset until a stopping condition is satisfied, such as the formation of a pure node, the attainment of a predefined maximum depth, or the availability of fewer than a specified number of samples. The terminal nodes represent the final predicted classes, and a generic procedure for constructing such a tree is shown in \cref{fig:tree-workflow}. Because the resulting hierarchy depends on how candidate splits are evaluated, different decision tree algorithms employ different selection criteria. For a dataset $D$ containing $K$ classes, let $p_k$ denote the proportion of samples belonging to class $k$. The entropy of $D$ is defined as

\begin{equation}\label{eq:entropy}
H(D)
=
-\sum_{k=1}^{K}p_k\log_2p_k.
\end{equation}

Entropy measures the uncertainty of the class distribution. A lower entropy value indicates that the samples within a node are more concentrated in a small number of classes, while an entropy of zero indicates a completely pure node. If an attribute $A$ partitions $D$ into $V$ subsets $D_v$, the corresponding information gain is

\begin{equation}\label{eq:gain}
\operatorname{Gain}(D,A)
=
H(D)
-
\sum_{v=1}^{V}
\frac{|D_v|}{|D|}
H(D_v).
\end{equation}

Information gain measures the reduction in uncertainty produced by a candidate split. ID3-style decision trees select the attribute with the highest information gain. However, the CART model used in this study constructs binary trees and evaluates candidate splits using Gini impurity:

\begin{equation}\label{eq:gini}
\operatorname{Gini}(D)
=
1-\sum_{k=1}^{K}p_k^2.
\end{equation}

The Gini impurity represents the probability of incorrectly labeling a randomly selected sample if its label were assigned according to the class distribution of the node, with a value of zero indicating that all samples belong to the same class. For each candidate split, CART calculates the weighted Gini impurity of the resulting child nodes and selects the feature and threshold that minimize this value. However, an unrestricted decision tree may continue splitting until it closely fits the training data, leading to excessive model complexity and poor generalization. To mitigate this risk, CART can apply cost-complexity pruning by minimizing

\begin{equation}\label{eq:pruning}
R_{\alpha}(T)
=
R(T)+\alpha|T|,
\end{equation}

where $R(T)$ is the empirical error of tree $T$, $|T|$ is the number of terminal nodes, and $\alpha$ is the complexity-control parameter. A smaller value of $\alpha$ retains a larger and more detailed tree, whereas a larger value penalizes additional terminal nodes and produces a simpler model. This trade-off allows the model to preserve useful decision boundaries while reducing overfitting, which is particularly important for VPID, where decision trees are well suited because they provide low-cost inference and explicit decision paths. In the vulnerability-prioritization module, each predicted priority can therefore be traced to a sequence of conditions involving factors such as CVSS score, asset value, network exposure, and exploitation history, thereby supporting human interpretation and manual review.

\subsection{Traffic features and containerization}
\label{subsec:traffic-tech}

Scapy provides programmable packet capture, dissection, and construction capabilities \cite{scapy2024usage}. In the proposed system, captured packets are grouped into sessions and converted into feature vectors containing traffic statistics, protocol structure information, and application layer content when available. For sessions containing plaintext payloads, textual content is represented using TF--IDF features based on established term weighting principles \cite{salton1988term}; for encrypted traffic, the system relies on flow metadata and statistical characteristics without inspecting application payloads~\citep{lin2022bert}. To simplify deployment, the capture and analysis components are packaged together with the scanning, storage, and presentation services in Docker containers, providing a consistent runtime environment while reducing deployment and maintenance effort. The detailed container architecture is presented in the implementation section.

\section{System Architecture}\label{sec:design}

\subsection{Requirements}\label{subsec:requirements}

The system manages internal assets and scan histories, invokes OpenVAS, ranks vulnerabilities with support for manual adjustment, and generates validation payloads for findings assessed as high risk. Its defense path captures traffic with Scapy, identifies attacks using machine learning, updates detection rules, verifies suspicious traffic with Snort, and blocks confirmed malicious sources. A web console provides authentication, dashboards, task and configuration management, operational logs, access control based on user roles, and audit trails. The system is designed to remain stable on hardware with limited resources, maintain low detection latency and a low false positive rate~\citep{yang2024lightweight}, support common network modes, and permit offline installation. To organize these functions, the VPID architecture consists of five layers. The data acquisition layer collects OpenVAS findings, Scapy traffic, and operational logs; the processing layer parses vulnerability data, cleans records, and constructs features; the analysis layer ranks vulnerabilities using a decision tree, detects attacks using Naive Bayes together with a preliminary decision tree filter, and selects validation payload templates; the defense layer performs rule verification, blocking, and alerting through Snort and iptables; and the presentation layer provides the web console, RESTful API, and notifications. These layers are implemented through a vulnerability module covering asset management, scanning, ranking, payload generation, and controlled validation, and a defense module covering traffic capture, parsing, feature extraction, model inference, rule verification, response, and alerting. Shared data management stores operational records, models, rules, and logs, while configuration management supplies common parameters for system operation, vulnerability scanning, and attack detection.

\subsection{Five-layer architecture}\label{subsec:architecture}

\begin{figure}[H]
  \centering
  \includegraphics[width=0.74\textwidth]{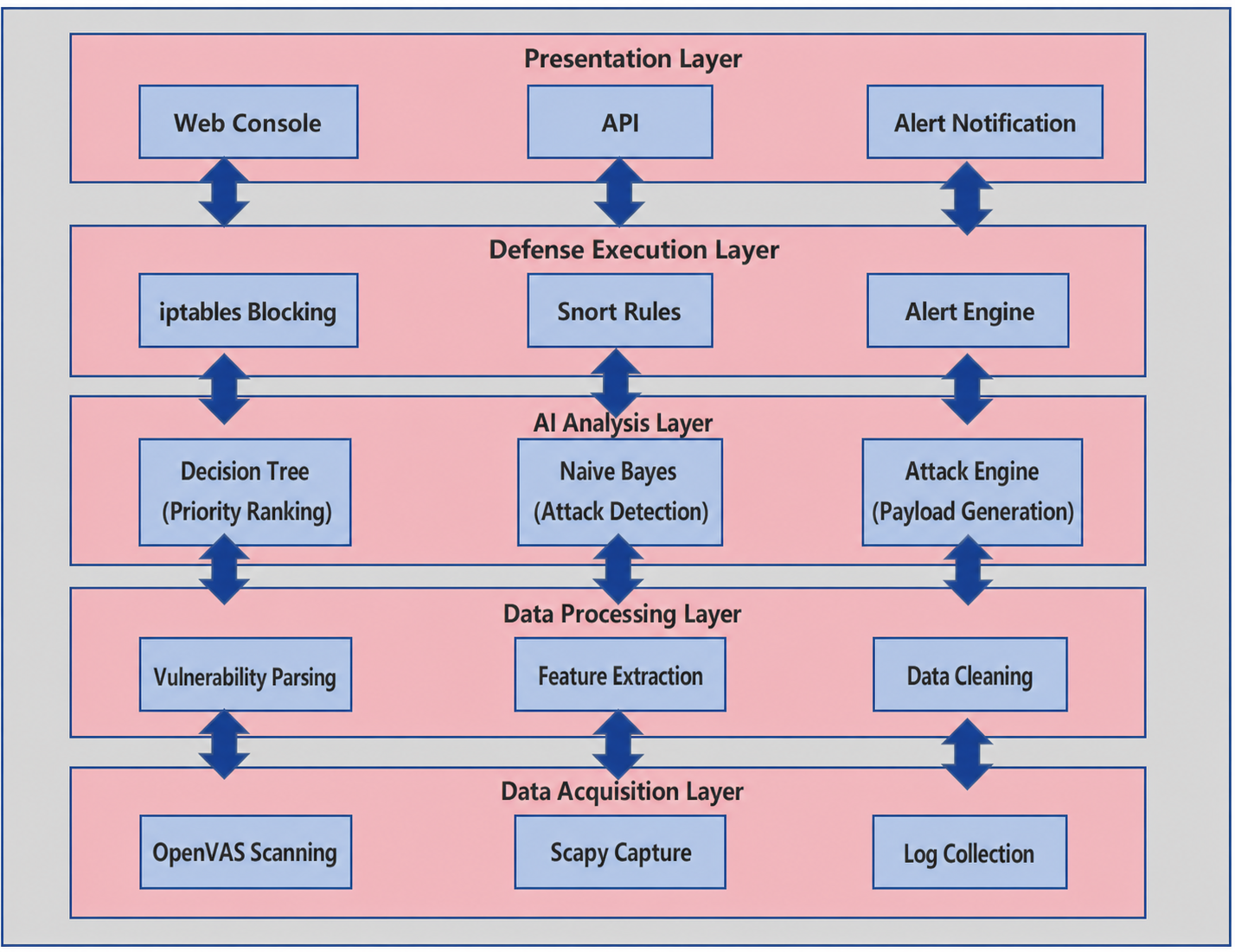}
  \caption{Five-layer system architecture.}
  \label{fig:architecture}
\end{figure}

The VPID architecture separates data acquisition, processing, machine-learning analysis, defense execution, and presentation layers, as shown in \cref{fig:architecture}. Components communicate through standardized interfaces so that scanners, models, rules, and storage can evolve independently.

\section{Vulnerability Prioritization and Validation}\label{sec:vulnerability}

\subsection{Processing pipeline}\label{subsec:vuln-pipeline}

\begin{figure}[!ht]
  \centering
  \includegraphics[
    width=0.75\linewidth,
    height=0.5\textheight,
    keepaspectratio,
    trim=30 30 30 30,
    clip
  ]{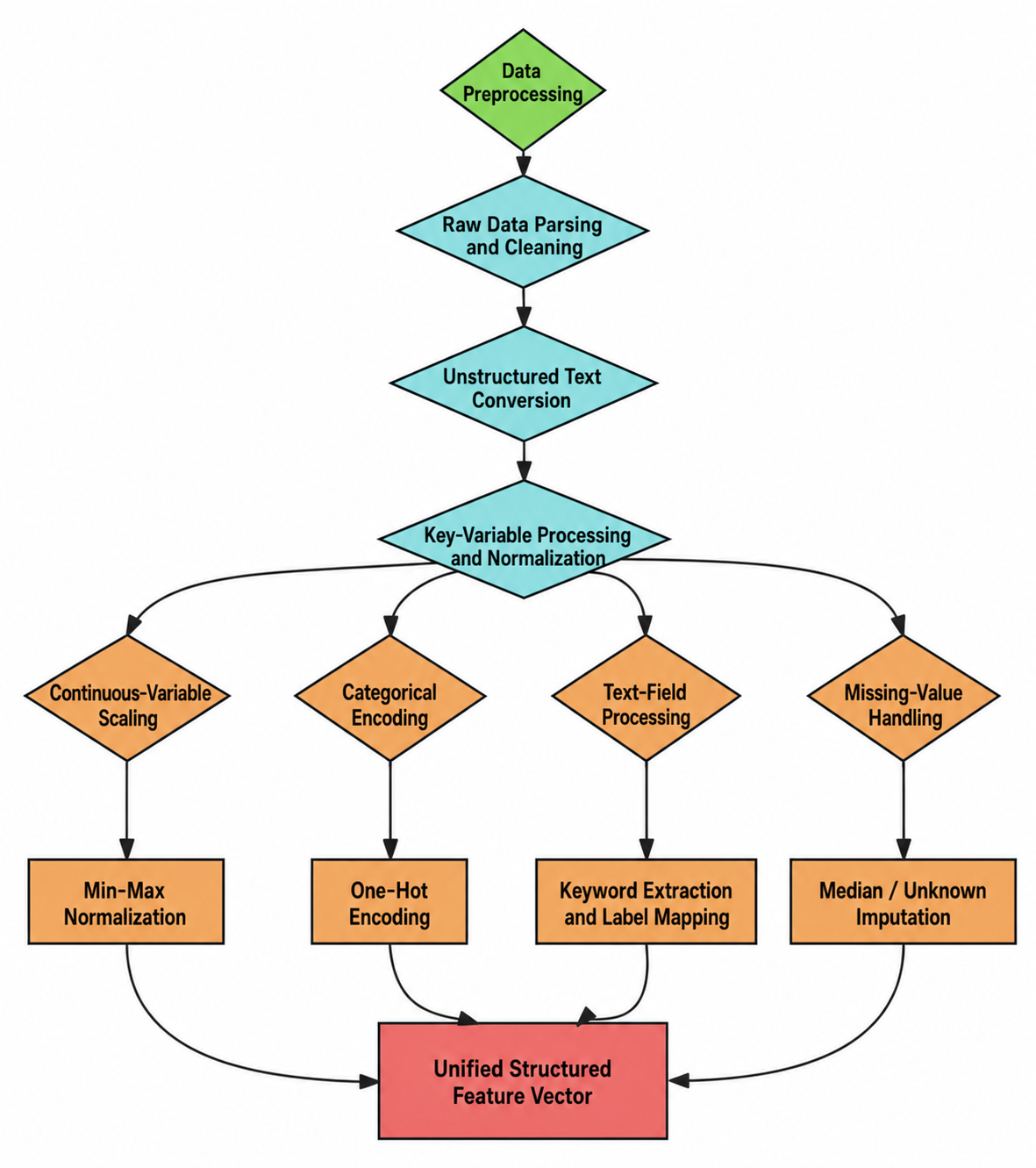}
  \caption{Vulnerability-record preprocessing and feature construction.}
  \label{fig:preprocessing}
\end{figure}

The vulnerability-prioritization module of VPID implements a traceable ``discover--assess--validate'' workflow. It reads the asset inventory, creates OpenVAS tasks, polls their status, retrieves XML reports, removes duplicates and invalid fields, normalizes fields, estimates priority, generates validation rules, executes controlled checks, and writes the result back. Each record contains the vulnerability name, target IP address, port, CVE identifier, CVSS score, description, remediation recommendation, and scanner-plugin identifier.
Continuous variables such as CVSS score, exposure count, and historical exploitation frequency are normalized by
\begin{equation}\label{eq:minmax}
x'=\frac{x-x_{\min}}{x_{\max}-x_{\min}}.
\end{equation}
Categorical properties such as vulnerability type, asset tier, and service type are one-hot encoded. Descriptions are mapped through keyword extraction, numerical missing values are filled with the median of the corresponding class, and categorical missing values use the literal category \texttt{Unknown}. These choices follow common intrusion-data preprocessing and feature-selection practice \cite{otokwala2024optimized,ayad2024hybrid,zhou2025hidim}. The complete flow is shown in \cref{fig:preprocessing}.

\subsection{Priority model}\label{subsec:priority-model}

Priority is formulated as a high/medium/low classification task. The CART model receives a structured feature vector built from CVSS, exploitability, asset and business value~\citep{yin2023empowering}, exposure, patch status, exploitation history, confidence, and the encoded categorical fields described above. At each node, the split minimizes the weighted Gini index
\begin{equation}\label{eq:gini-index}
\operatorname{GiniIndex}(D,a)=\sum_{v}\frac{|D_v|}{|D|}\operatorname{Gini}(D_v),
\end{equation}
and cost-complexity pruning uses
\begin{equation}\label{eq:cart-cost}
C_{\alpha}(T)=C(T)+\alpha|T|.
\end{equation}
The source algorithm is preserved in \cref{alg:vulnerability-priority}.

\begin{algorithm}[htbp]
\caption{Vulnerability-priority assessment}\label{alg:vulnerability-priority}
\begin{algorithmic}[1]
\Require vulnerability set $V$, feature matrix $X$, decision tree $T$
\Ensure priority labels and comprehensive risk scores
\For{each vulnerability $v_i\in V$}
  \State $x_i\gets\{\textit{cvss},\textit{exploitability},\textit{assetValue},\textit{businessValue},$
  \Statex \hspace{2.7em}$\textit{exposure},\textit{patch},\textit{history},\textit{confidence}\}$
  \State $p_i\gets\Call{TreePredict}{T,x_i}$
  \State $s_i\gets w_1\textit{cvss}+w_2\textit{exploitability}+w_3\textit{assetValue}$
  \Statex \hspace{2.7em}$+w_4\textit{businessValue}+w_5\textit{exposure}+w_6\textit{history}-w_7\textit{patch}+w_8\textit{confidence}$
  \If{$p_i\ge\theta_{\mathrm{high}}$ and $s_i\ge\tau_{\mathrm{high}}$}
    \State $\textit{Priority}_i\gets\mathrm{High}$
  \ElsIf{$p_i\ge\theta_{\mathrm{mid}}$ and $s_i\ge\tau_{\mathrm{mid}}$}
    \State $\textit{Priority}_i\gets\mathrm{Medium}$
  \Else
    \State $\textit{Priority}_i\gets\mathrm{Low}$
  \EndIf
\EndFor
\State sort vulnerabilities by $s_i$ in descending order
\State \Return $(\textit{Priority},s)$
\end{algorithmic}
\end{algorithm}

\subsection{Payload rule engine}\label{subsec:payload-engine}

JSON rules store parameterized validation templates for SQL injection, cross-site scripting, command injection, and file inclusion. For a high-priority finding, the engine substitutes the target IP address, port, path, parameter position, and encoding, then applies escaping and retry logic. Candidate template $r_i$ is ranked by
\begin{equation}\label{eq:payload-score}
\operatorname{Score}(r_i)=w_1M_{\mathrm{type}}+w_2M_{\mathrm{port}}+w_3M_{\mathrm{param}}+w_4M_{\mathrm{keyword}},
\end{equation}
where each $M$ is its corresponding match result and the empirical weights satisfy $\sum_{j=1}^{4}w_j=1$. The highest-scoring rule produces a validation payload in a controlled environment; the module records rather than assumes exploitability.

\section{Attack Detection and Response}\label{sec:defense}

\subsection{Capture, reconstruction, and representation}
\label{subsec:defense-input}

The attack-detection module processes both online and offline traffic. Online data consist of packets captured from selected network interfaces and are used for real-time detection and response, whereas offline data consist of locally collected traffic and public attack samples and are used for model training, parameter selection, and independent evaluation. Although the two data sources are obtained in different ways, they are processed using the same session-reconstruction and feature-extraction procedure to maintain consistency between model training and online inference. Specifically, for each captured packet, the module records the source and destination addresses, source and destination ports, transport protocol, packet length, and timestamp. When unencrypted application content is available, the corresponding request method, path, parameters, status information, and payload text are also extracted. HTTPS payloads are not decrypted by the module; therefore, encrypted sessions contribute only flow metadata, protocol information, and statistical characteristics~\citep{han2024gnn}.

Before session reconstruction, duplicate and malformed packets are removed, while TCP control packets without application data are retained for flow statistics but excluded from payload analysis. The remaining packets are ordered by timestamp and grouped using a normalized tuple containing the two endpoints, their ports, and the transport protocol. Endpoint order is normalized so that packets travelling in opposite directions are assigned to the same bidirectional flow. A new session is created when the interval between consecutive packets exceeds the inactivity threshold $\tau$. When application content is visible, payload preprocessing normalizes URL encoding, letter case, escape characters, repeated delimiters, and parameter separators, thereby reducing superficial differences between semantically similar requests and limiting simple encoding based evasion. Formally, session $k$ is defined as

\begin{equation}\label{eq:session}
\operatorname{Session}_k
=
\left\{
p_j
\;\middle|\;
\operatorname{key}(p_j)=\operatorname{key}_k,\,
t_j-t_{j-1}<\tau
\right\}.
\end{equation}

Here, $\operatorname{key}(p_j)$ denotes the normalized five-tuple of packet $p_j$, and $t_j$ denotes its capture time. The timeout parameter $\tau$ prevents unrelated requests with the same endpoints from being merged into a single long-lived session. Based on the reconstructed sessions, features that cannot be derived from isolated packets—such as connection frequency, request–response size relationships, inter-arrival patterns~\citep{chai2024combo}, and repeated access behavior—can be calculated. Each session is then represented using three feature groups. The first group comprises statistical and protocol metadata, including connection frequency, request length, packet count, parameter count, abnormal character ratio, request method, response status, URL depth, and parameter position. These features characterize session structure and behavior without relying entirely on payload content. The second group consists of TF--IDF features extracted from visible payload text, which assign greater weight to terms that occur frequently in a particular session but less frequently across the training corpus. For a term $t$, a session document $d$, a corpus containing $N$ documents, and $n_t$ documents containing $t$, the smoothed TF--IDF value is

\begin{equation}\label{eq:tfidf}
\operatorname{TFIDF}(t,d)
=
\operatorname{tf}(t,d)
\left(
\log\frac{1+N}{1+n_t}+1
\right).
\end{equation}

The smoothing terms prevent division by zero and ensure nonnegative feature values, which are required by the multinomial Naive Bayes classifier. Terms associated with injection operators, suspicious commands, traversal sequences, and script fragments can consequently receive different weights according to their distribution in the training corpus. To complement these token-level features, the third group contains character trigrams, which preserve partial attack indicators even when spacing, encoding, or token boundaries are altered, thereby capturing obfuscated or fragmented malicious strings that complete tokens may fail to retain. Together, these complementary feature groups are combined to form the final representation:

\begin{equation}\label{eq:joint-vector}
x
=
\left[
x_{\mathrm{meta}},
x_{\mathrm{tfidf}},
x_{\mathrm{3gram}}
\right],
\end{equation}

where $x_{\mathrm{meta}}$ contains the statistical and protocol metadata, while $x_{\mathrm{tfidf}}$ and $x_{\mathrm{3gram}}$ represent payload text. The experimental configuration uses 12 metadata dimensions, 100 TF--IDF dimensions, and 50 character-three-gram dimensions, giving a total of

\begin{equation}
\dim(x)=12+100+50=162.
\end{equation}

For two-stage detection, the metadata vector is supplied to the decision tree, whereas the combined text vector
$x_{\mathrm{text}}=[x_{\mathrm{tfidf}},x_{\mathrm{3gram}}]$
is supplied to multinomial Naive Bayes. This separation allows the first stage to process every session at low cost while reserving payload-oriented classification for sessions that require further analysis.

\subsection{Two-stage detector and response}
\label{subsec:mixed-model}

The intrusion-detection module of VPID adopts a two-stage architecture to balance computational cost and detection capability~\citep{li2024two,chen2024machine}. The first stage uses a decision tree to identify sessions that match obvious normal-traffic patterns. Because the tree operates primarily on compact metadata features, it can quickly process a large volume of sessions and discard low-risk traffic without running the text classifier on every input. The filtering threshold $\theta_{\mathrm{tree}}$ determines how much traffic is forwarded to the second stage. A lower threshold retains more ambiguous sessions for further analysis, improving sensitivity at the cost of additional computation.

For a session that passes the first-stage filter, multinomial Naive Bayes analyzes the TF--IDF and character-three-gram features. Let $\mathcal{C}$ denote the set of candidate classes. In the experiment, these classes include SQL injection, cross-site scripting, command injection, directory traversal, and other attacks. The classifier produces a posterior score for each class:

\begin{align}
q_c
&=
P(c\mid x_{\mathrm{text}}),
\qquad c\in\mathcal{C},
\label{eq:posterior}\\
\hat{c}
&=
\arg\max_{c\in\mathcal{C}}q_c,
\qquad
p_{\mathrm{nb}}
=
\max_{c\in\mathcal{C}}q_c.
\label{eq:posterior-decision}
\end{align}

Here, $\hat{c}$ is the predicted attack class, and $p_{\mathrm{nb}}$ is the corresponding classification confidence. Laplace smoothing prevents unseen features from producing zero class likelihoods and improves classifier stability when sparse payload representations are used. Because machine learning outputs are not used as the sole basis for response decisions, Snort compatible rules provide an additional signature based verification signal \cite{snort2024rules}. Let $p_{\mathrm{tree}}\in[0,1]$ denote the suspicious traffic score produced by the decision tree, and let $I_{\mathrm{rule}}\in{0,1}$ indicate whether a rule matches the reconstructed session. These three signals are combined into the risk score

\begin{equation}\label{eq:risk-fusion}
R
=
\alpha p_{\mathrm{tree}}
+
\beta p_{\mathrm{nb}}
+
\gamma I_{\mathrm{rule}},
\qquad
\alpha+\beta+\gamma=1,
\end{equation}

where $\alpha$, $\beta$, and $\gamma$ are nonnegative weights. The decision-tree score represents structural abnormality, the Naive Bayes score represents payload-level attack confidence, and the rule indicator represents explicit signature evidence. Combining these signals reduces dependence on a single model and allows strong rule evidence to increase the priority of an otherwise uncertain detection result~\citep{li2024hda}.

\begin{figure}[htbp]
  \centering
  \includegraphics[width=0.86\textwidth]{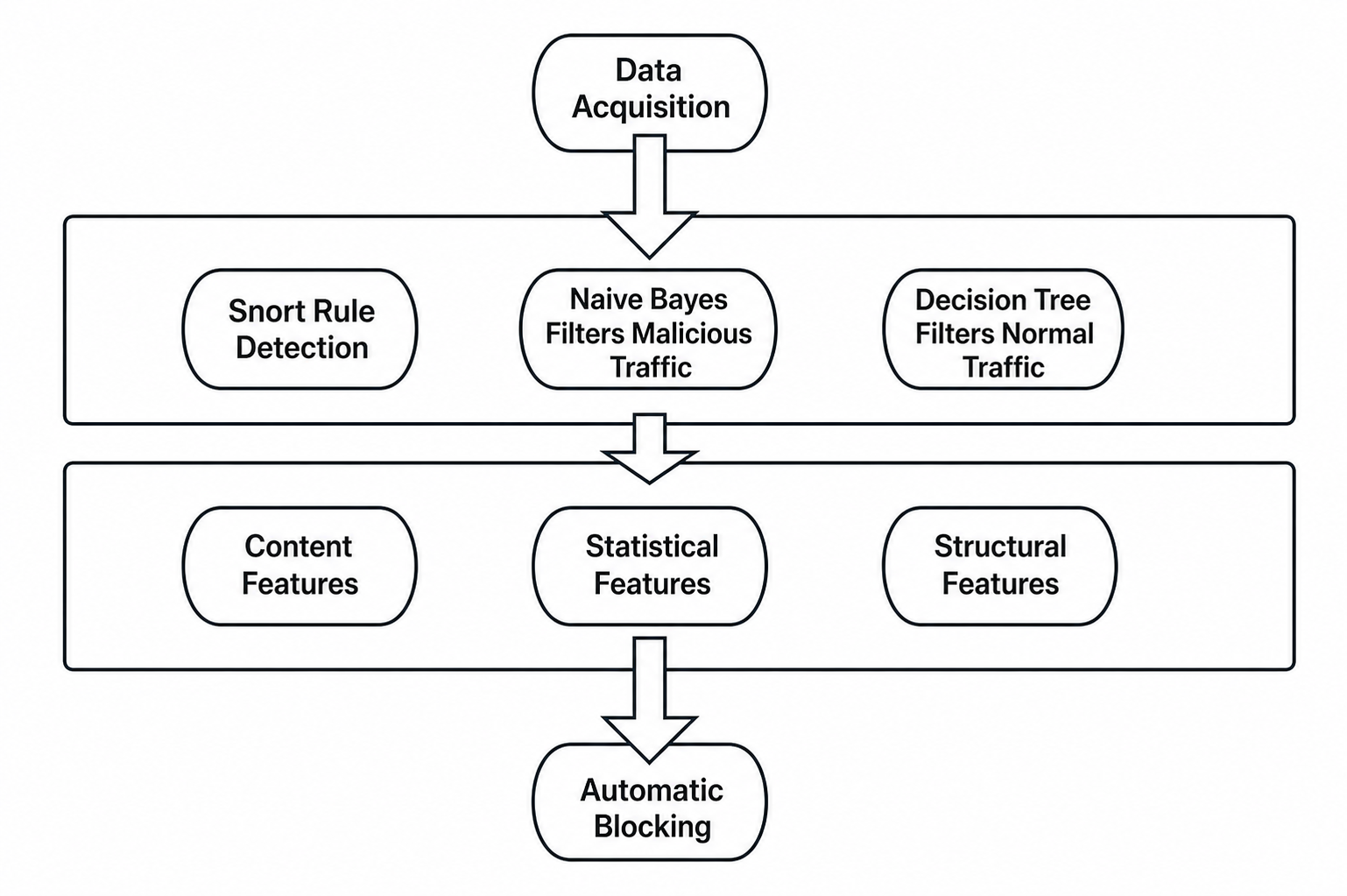}
  \caption{Defense-module data and decision flow.}
  \label{fig:defense-overview}
\end{figure}

Two response thresholds divide the risk score into three operational levels. When $R\geq\theta_{\mathrm{block}}$, the session is treated as high risk and its source is submitted to the iptables response component. When $\theta_{\mathrm{alert}}\leq R<\theta_{\mathrm{block}}$, the event is retained as a medium risk alert for manual review. When $R<\theta_{\mathrm{alert}}$, no active response is applied, but the decision is logged for later analysis. The thresholds satisfy $0\leq\theta_{\mathrm{alert}}<\theta_{\mathrm{block}}\leq1$. By separating classification, risk assessment, and response, the system prevents an attack label from automatically triggering a blocking action without considering classification confidence and rule evidence. This design also allows response thresholds and fusion weights to be adjusted for different operational environments without retraining the underlying classifiers. The complete data and decision flow is shown in \cref{fig:defense-overview}.

The online decision procedure is summarized in \cref{alg:mixed-detection}. The metadata vector is first evaluated by the decision tree. Sessions below the screening threshold are recorded as low-risk normal traffic. For the remaining sessions, the text classifier determines the most likely attack category, after which the model scores and rule-matching result are combined. The returned tuple contains the predicted class, risk level, and response action, thereby separating detection semantics from enforcement policy.

\begin{algorithm}[htbp]
\caption{Mixed-model traffic detection}
\label{alg:mixed-detection}
\begin{algorithmic}[1]
\Require session $S$; thresholds $\theta_{\mathrm{tree}},
\theta_{\mathrm{alert}},\theta_{\mathrm{block}}$;
weights $\alpha,\beta,\gamma$
\Ensure detection result, risk level, and action

\State extract $X_{\mathrm{meta}}$ and $X_{\mathrm{text}}$
\State $p_{\mathrm{tree}}\gets
\Call{TreePredict}{X_{\mathrm{meta}}}$

\If{$p_{\mathrm{tree}}<\theta_{\mathrm{tree}}$}
  \State \Return $(\mathrm{Normal},\mathrm{Low},\mathrm{Log})$
\EndIf

\State $q(c)\gets\Call{NBPredict}{X_{\mathrm{text}}}$
\State $\textit{Result}\gets\arg\max_c q(c)$
\State $p_{\mathrm{nb}}\gets\max_c q(c)$
\State $I_{\mathrm{rule}}\gets\Call{RuleMatch}{S}$

\State $R\gets
\alpha p_{\mathrm{tree}}
+\beta p_{\mathrm{nb}}
+\gamma I_{\mathrm{rule}}$

\If{$R\geq\theta_{\mathrm{block}}$}
  \State \Return $(\textit{Result},\mathrm{High},\mathrm{Block})$
\ElsIf{$R\geq\theta_{\mathrm{alert}}$}
  \State \Return $(\textit{Result},\mathrm{Medium},\mathrm{Alert})$
\Else
  \State \Return $(\textit{Result},\mathrm{Low},\mathrm{Log})$
\EndIf
\end{algorithmic}
\end{algorithm}

\begin{figure}[htbp]
  \centering
  \includegraphics[width=0.9\textwidth]{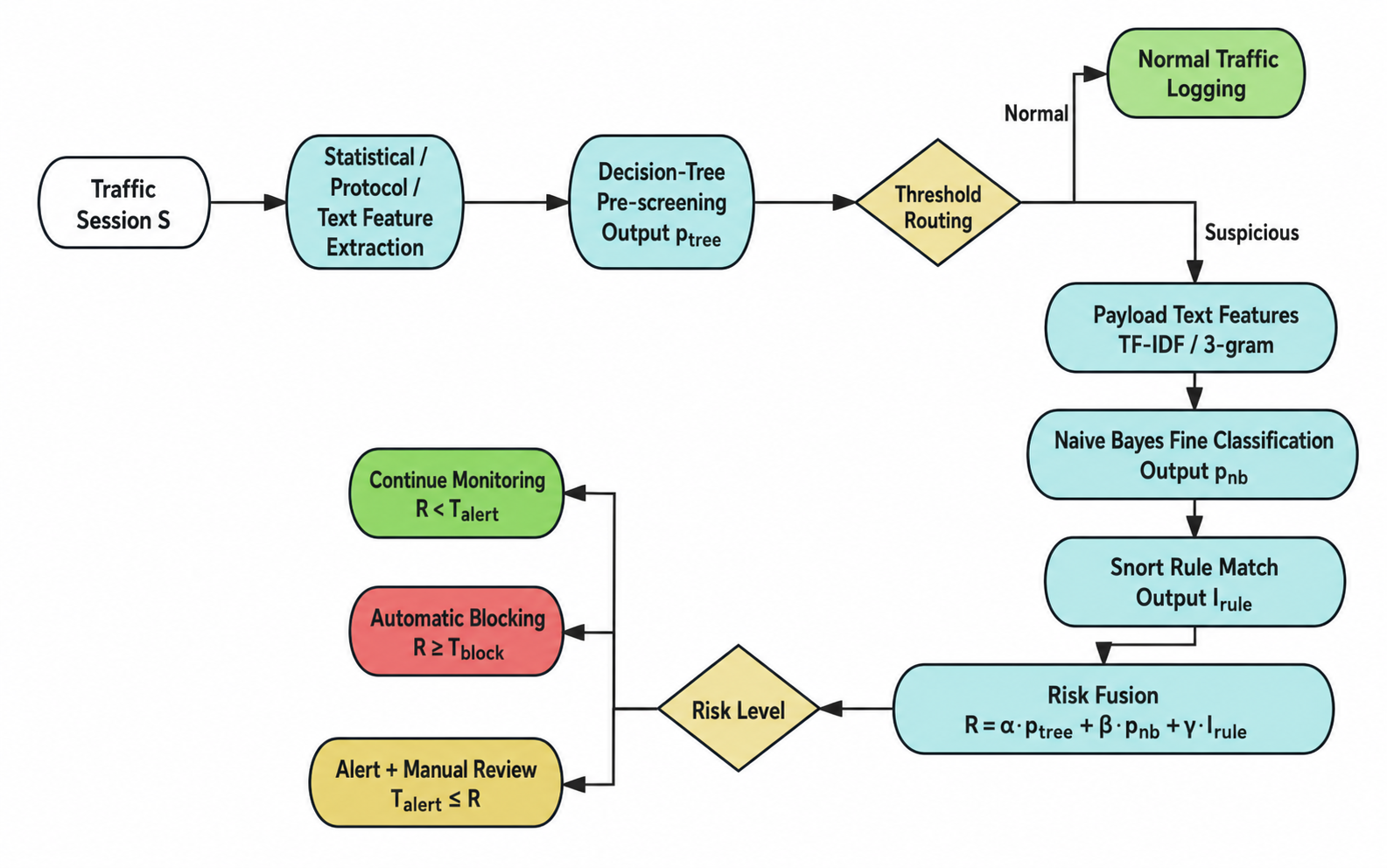}
  \caption{Mixed decision-tree, Naive Bayes, and Snort detection workflow.}
  \label{fig:mixed-workflow}
\end{figure}

The two stage structure and the interaction among machine learning inference, rule verification, and response enforcement are illustrated in \cref{fig:mixed-workflow}. The decision tree provides low cost screening, multinomial Naive Bayes identifies attack categories for suspicious sessions, and Snort rules supply explicit evidence of known attack patterns. The fused risk score is then mapped to logging, alerting, or blocking. For traceability and subsequent analysis, the system records the model outputs, rule matching result, risk score, response action, and manual review outcome for each session. These records help explain response decisions, examine false positives, adjust thresholds, and update the model and rule repositories during offline review~\citep{sharafaldin2018toward}.

\FloatBarrier
\section{Implementation and Deployment}\label{sec:implementation}

The development and test host is equipped with two CPU cores, 4~GB of memory, and a 40-GB disk. The core software stack is built on Python~3.9, with scikit-learn implementing the two machine-learning models, MySQL managing assets, vulnerabilities, attacks, and configuration data, and a lightweight Web framework providing the user interface. Python interacts with OpenVAS through \texttt{gvm-tools}, while Scapy is responsible for packet capture and the host's \texttt{iptables} rule chain enforces blocking policies.The VPID framework is deployed and managed through Docker Compose, which starts and coordinates the required services, while external volumes ensure the persistence of logs, models, and business data across container recreation.

The experimental environment uses a dedicated internal network topology comprising a target host with known vulnerable Web applications, a defense server running the four services, and a separate test terminal. Controlled scans, validation requests, and attack traffic are issued from the test terminal, allowing the complete process from asset discovery and controlled vulnerability validation to traffic detection and response to be observed. Within this environment, model calibration evaluates the smoothing parameter of multinomial Naive Bayes over the set

\[
\alpha\in\{0.1,0.3,0.5,0.7,1.0\}.
\]
The source experiment selects $\alpha=1.0$; all reported values are retained in \cref{tab:smoothing}.

\begin{table}[htbp]
\centering
\caption{Effect of the Naive-Bayes smoothing factor.}
\label{tab:smoothing}
\begin{tabular}{@{}ccccc@{}}
\toprule
$\alpha$ & Precision & Recall & F1 score & FPR \\
\midrule
0.1 & 0.921 & 0.869 & 0.894 & 0.023 \\
0.3 & 0.926 & 0.876 & 0.900 & 0.021 \\
0.5 & 0.930 & 0.881 & 0.903 & 0.018 \\
0.7 & 0.933 & 0.885 & 0.905 & 0.014 \\
1.0 & 0.945 & 0.883 & 0.913 & 0.012 \\
\bottomrule
\end{tabular}
\end{table}

\section{Evaluation}\label{sec:experiments}

\subsection{Datasets and feature mapping}\label{subsec:datasets}

The traffic corpus combines CICIDS2017~\citep{sharafaldin2018toward}, NSL-KDD~\citep{tavallaee2009detailed}, and UNSW-NB15~\citep{moustafa2015unsw} with locally collected traffic. These sources differ in collection environment, record format, feature definitions, and label granularity~\citep{bonninghausen2024introducing}. Therefore, they are first mapped to a common session-level representation before model training~\citep{goldschmidt2025network}. Timestamps, transport protocols, addresses, ports, service identifiers, and available labels are normalized into a unified schema. Empty and duplicate records are removed, while control-only traffic is excluded when packet-level information is available.

The experiment retains Web related records identified through ports 80 and 443 or the corresponding service labels. Payload features are generated only when application layer content is visible; encrypted HTTPS samples and records without visible payloads contribute only metadata and statistical features rather than artificially reconstructed text. Original labels are mapped to normal, SQL injection, cross site scripting (XSS), command injection, directory traversal, and other attacks. A sample is assigned to a specific Web attack class only when supported by its source annotation, while labels that cannot be mapped reliably are placed in the other attacks category. This procedure avoids inferring fine grained attack types from incomplete labels and ensures that the dataset and features match the intended detection task \cite{pinto2025review}. After preprocessing, the training corpus contains 550,000 samples, with the class distribution reported in \cref{tab:traffic-dataset}.

\begin{table}[htbp]
\centering
\caption{Traffic training-dataset composition.}
\label{tab:traffic-dataset}
\begin{tabular}{@{}lrr@{}}
\toprule
Class & Samples & Share \\
\midrule
Normal traffic & 450,000 & 81.8\% \\
SQL injection & 35,000 & 6.4\% \\
XSS & 28,000 & 5.1\% \\
Command injection & 15,000 & 2.7\% \\
Directory traversal & 12,000 & 2.2\% \\
Other attacks & 10,000 & 1.8\% \\
\midrule
Total & 550,000 & 100\% \\
\bottomrule
\end{tabular}
\end{table}

Normal traffic accounts for 81.8\% of the training corpus, while all attack classes together account for 18.2\%. SQL injection and XSS are the largest attack categories, whereas command injection, directory traversal, and other attacks occur less frequently~\citep{abdelkhalek2023addressing}. This distribution reflects the predominance of normal traffic in operational networks but also makes overall accuracy insufficient for evaluation; therefore, precision, recall, F1 score, and false positive rate are reported separately. For model input, each sample is transformed into a 162 dimensional vector comprising 12 statistical and protocol metadata dimensions, 100 TF--IDF payload dimensions, and 50 character trigram dimensions. The vocabulary and feature mapping are fitted on the training data and applied unchanged to the independent test set, which contains 55,000 samples and is excluded from model fitting and parameter selection.

For the vulnerability prioritization experiment, 15,000 historical vulnerability scan records are used, containing CVSS information together with asset, exposure, patch, exploitation history, and confidence features. Their recorded priorities are mapped to High, Medium, and Low classes, as shown in \cref{tab:vulnerability-dataset}. This dataset is less imbalanced than the traffic corpus, although Medium priority records still represent half of all samples. The decision tree, random forest, and CVSS only baseline are evaluated using identical training and test partitions, ensuring that performance differences result from the models rather than different sample selections.

\begin{table}[htbp]
\centering
\caption{Vulnerability-priority dataset.}
\label{tab:vulnerability-dataset}
\begin{tabular}{@{}lrr@{}}
\toprule
Priority & Samples & Share \\
\midrule
High & 4,500 & 30\% \\
Medium & 7,500 & 50\% \\
Low & 3,000 & 20\% \\
\midrule
Total & 15,000 & 100\% \\
\bottomrule
\end{tabular}
\end{table}

\subsection{Metrics}\label{subsec:metrics}

The experiments report precision, recall, F1 score, and false positive rate to evaluate different operational consequences. For attack detection, attack traffic is treated as the positive class and normal traffic as the negative class. Precision reflects the reliability of predicted attacks, recall reflects the ability to detect actual attacks, and the false positive rate measures the proportion of normal samples incorrectly classified as attacks. For vulnerability ranking, these metrics are calculated separately for the High, Medium, and Low classes using a one versus rest formulation and then averaged across the three classes. Accordingly, the reported vulnerability prioritization value of 91.8\% and attack detection value of 94.5\% refer to precision rather than accuracy. Low precision leads to unnecessary alerts or blocking actions, whereas low recall leaves more attacks undetected; the F1 score balances these concerns, while the false positive rate directly indicates the impact on benign traffic. The metrics are calculated as follows:

\subsection{Vulnerability ranking}\label{subsec:vuln-results}

The CART decision tree is compared with a random forest and a CVSS only baseline using the same data partition. The CVSS only baseline uses the CVSS score as its sole predictive input, whereas the two machine learning models additionally incorporate contextual features such as asset value, exposure, patch status, and exploitation history~\citep{iannone2024early}. This comparison evaluates both the contribution of contextual information and the tradeoff between a single interpretable tree and an ensemble model. The results are shown in \cref{tab:vulnerability-results}.

\begin{table}[htbp]
\centering
\caption{Vulnerability-priority ranking results.}
\label{tab:vulnerability-results}
\begin{tabular}{@{}lccc@{}}
\toprule
Model & Precision & Recall & F1 score \\
\midrule
Decision tree & 0.918 & 0.895 & 0.906 \\
Random forest & 0.932 & 0.908 & 0.920 \\
CVSS score only & 0.756 & 0.742 & 0.749 \\
\bottomrule
\end{tabular}
\end{table}

The random forest produces the highest scores, exceeding the decision tree by 1.4 percentage points in precision and F1 score and by 1.3 percentage points in recall. However, the decision tree still achieves an F1 score of 90.6\% and provides an explicit decision path for each predicted priority.

More importantly, the decision tree improves the F1 score over the CVSS-only baseline by 15.7 percentage points. Its precision and recall are also 16.2 and 15.3 percentage points higher, respectively. This difference indicates that vulnerability severity alone is insufficient for priority ranking and that asset, exposure, patch, and exploitation-related features contribute useful information~\citep{xu2025exploitability}. Given the relatively small performance gap from the random forest, the system selects the decision tree to preserve direct decision-path explanations and simpler model inspection.

\subsection{Attack detection}\label{subsec:attack-results}

The VPID framework, which combines multinomial Naive Bayes with decision-tree screening, is compared with SVM, random forest, and standalone Snort rules. To ensure a consistent comparison, all machine learning models use the same training and test partitions, while the Snort baseline relies solely on signature matching without model confidence fusion. This comparison examines whether the proposed two stage framework improves attack detection coverage while maintaining an acceptable false positive rate. The aggregate results are shown in \cref{tab:attack-results}.

\begin{table}[htbp]
\centering
\caption{Attack detection results.}
\label{tab:attack-results}
\begin{tabular}{@{}lccc@{}}
\toprule
Model & Precision & Recall & F1 score \\
\midrule
Naive Bayes + decision tree & 0.945 & 0.883 & 0.913 \\
SVM & 0.941 & 0.876 & 0.907 \\
Random forest & 0.947 & 0.860 & 0.901 \\
Snort rules & 0.882 & 0.795 & 0.836 \\
\bottomrule
\end{tabular}
\end{table}

The random forest achieves the highest precision at 94.7\%, exceeding the proposed detector by 0.2 percentage points, but its recall is 2.3 percentage points lower, resulting in an F1 score 1.2 percentage points below that of the proposed detector. Compared with SVM, the VPID detector improves precision, recall, and F1 score by 0.4, 0.7, and 0.6 percentage points, respectively. Standalone Snort rules achieve the lowest recall because signature matching detects only attacks covered by existing rules; the proposed detector improves precision and recall over Snort by 6.3 and 8.8 percentage points, respectively. These results indicate that combining statistical screening and payload classification with rule based verification provides a better balance than relying solely on either machine learning or fixed signatures. And the independent 55,000 sample confusion matrix is reported in \cref{tab:confusion}.

\begin{table}[htbp]
\centering
\caption{Attack-detection confusion matrix.}
\label{tab:confusion}
\begin{tabular}{@{}lrr@{}}
\toprule
 & Predicted normal & Predicted attack \\
\midrule
Actual normal & 44,285 & 523 \\
Actual attack & 1,191 & 9,001 \\
\bottomrule
\end{tabular}
\end{table}

The test set contains 44,808 normal samples and 10,192 attack samples, of which 523 normal samples are incorrectly classified as attacks and 1,191 attacks are missed. These results correspond to a precision of 94.5\%, a recall of 88.3\%, and a false positive rate of 1.2\%. Although the low false positive rate limits unnecessary alerts from benign traffic, the 1,191 false negatives indicate that the detector cannot identify every attack. This limitation supports retaining Snort verification, audit records, and manual review rather than treating the classifier as an infallible blocking mechanism.

\subsection{Response time and resource use}\label{subsec:performance}

The performance evaluation considers both user-facing operations and security-processing latency. Web page load time is used to measure the responsiveness of the management interface, while attack-detection latency covers the complete process of feature-vector generation and model inference. Defense-response time measures the interval from attack detection to the generation of the corresponding response action, whereas vulnerability scanning is evaluated separately because it is a batch operation whose duration depends on factors such as the number of targets, available services, and enabled vulnerability tests. All measurements were conducted on the two-core, 4-GB host described above, and the results are reported in \cref{tab:response-time}. Under this environment, the mean model-detection latency is 65 ms, with a maximum observed latency of 120 ms. After rule verification and response enforcement are included, the mean end-to-end defense time increases to 280 ms but remains below 500 ms in the reported measurements. In contrast, scanning 100 IP addresses requires an average of 28 minutes because OpenVAS performs a substantially larger set of network and service checks. Despite these processing requirements, CPU utilization remains below 15\% and memory consumption is approximately 800MB during the reported evaluation workload, indicating that the online detection and response path can operate within the intended two-core, 4-GB deployment environment.

\begin{table}[htbp]
\centering
\caption{System response times.}
\label{tab:response-time}
\begin{tabular}{@{}lrr@{}}
\toprule
Operation & Mean & Maximum \\
\midrule
Web page load & 1.2 s & 2.8 s \\
Vulnerability scan (100 IP addresses) & 28 min & 35 min \\
Attack-detection latency & 65 ms & 120 ms \\
Defense-response time & 280 ms & 450 ms \\
\bottomrule
\end{tabular}
\end{table}

\subsection{Limitations}\label{subsec:limitations}

The evaluation and implementation have several limitations. First, encrypted traffic is represented using statistical and protocol metadata, so payload-based attacks inside HTTPS sessions may not be visible without authorized decryption. Second, the vulnerability workflow depends on the OpenVAS vulnerability-test feed and cannot reliably identify vulnerabilities for which no applicable test or signature exists. Third, combining datasets collected in different environments may introduce distribution differences even after field and label alignment; consequently, performance on the constructed corpus may not fully represent every production network. Fourth, the learned detector may be bypassed by adversarially modified payloads or traffic patterns that differ substantially from the training distribution. Finally, the centralized architecture is designed for small-enterprise networks and may encounter capture, storage, or inference bottlenecks under substantially larger traffic volumes.

\FloatBarrier
\section{Conclusion and Future Work}\label{sec:conclusion}

This study presents VPID, an integrated vulnerability prioritization and intrusion detection framework for resource-constrained small-enterprise networks. Rather than treating vulnerability management and intrusion detection as isolated tasks, the VPID framework connects vulnerability discovery, context-aware prioritization, controlled validation, traffic detection, rule-based verification, and automated response through a unified workflow, whose effectiveness and practicality are demonstrated from three aspects. First, the vulnerability-management workflow integrates OpenVAS scanning, CART-based priority classification, and rule-based controlled validation. Evaluation on 15,000 labeled vulnerability records achieved 91.8\% precision, 89.5\% recall, and an $F_1$ score of 90.6\%, indicating that contextual features provide more informative prioritization results than CVSS scores alone. Second, the intrusion-detection and response pipeline combines statistical and protocol metadata, TF--IDF and character-three-gram features, decision-tree screening, multinomial Naive Bayes classification, Snort-based verification, and \texttt{iptables} enforcement. Based on 550,000 training samples and an independent test set of 55,000 samples, the detector achieved 94.5\% precision, 88.3\% recall, and an $F_1$ score of 91.3\%, with a false-positive rate of approximately 1.2\%. Finally, the complete system adopts a five-layer architecture and is deployed through four coordinated Docker services responsible for scanning, analysis, persistent storage, and user interaction. Under the evaluated workload, the system operated on a two-core, 4-GB host with CPU utilization below 15\% and memory consumption of approximately 800MB, demonstrating its feasibility for lightweight security-management scenarios.

These results demonstrate that lightweight and interpretable models can support an integrated security workflow without requiring high-end computing resources. The system maintains explicit decision paths, rule-matching evidence, and response records, enabling automated results to be inspected and reviewed by operators. However, the current evaluation is limited by the visibility of encrypted traffic, the coverage of the OpenVAS vulnerability-test feed, possible distribution differences among the combined datasets, and the throughput constraints of centralized deployment. Future work will investigate encrypted-traffic analysis based on TLS fingerprints~\citep{li2025hierarchical}, packet-size sequences, and temporal flow characteristics without direct payload inspection, while authorized traffic decryption may be explored in controlled internal environments where policies and regulations permit. In addition, robustness and generalization will be improved through temporally separated evaluation data, adversarially modified traffic samples~\citep{nitish2024class}, adversarial training, and input-normalization strategies. Lightweight model ensembles and distributed traffic collection will also be studied to enhance detection coverage and scalability while preserving the low-resource deployment objective of the system.
\section{Acknowledgments}
AI-based tools are used for language polishing during manuscript preparation.

\nocite{*}

\begingroup
\small
\bibliographystyle{plain}
\bibliography{references}
\endgroup

\end{document}